\documentclass[12pt]{article}
\usepackage{amsmath,amssymb}
\usepackage{epsfig,euscript,array,cite}
\usepackage{amssymb}
\usepackage{graphics}
\usepackage{graphicx}

\newcommand{\startappendix}{
\setcounter{section}{0}
\renewcommand{\thesection}{\Alph{section}}}

\newcommand{\Appendix}[1]{
\refstepcounter{section}
\begin{flushleft}
{\large\bf Appendix \thesection: #1}
\end{flushleft}}

\newcommand{\cN}{{\cal N}}

\newcommand{\cL}{{\cal L}}

\def\a{\alpha}
\def\b{\beta}

\def\G{\Gamma}
\def\D{\Delta}

\def\e{\varepsilon}
\def\t{\theta}

\def\k{\kappa}

\def\s{\sigma}

\def\beq{\begin{equation}}
\def\eeq{\end{equation}}
\def\beqn{\begin{eqnarray}}
\def\eeqn{\end{eqnarray}}
\def\ba{\begin{eqnarray}}
\def\ea{\end{eqnarray}}

\def\l{\langle}

\def\xprim2bar{\overline{x}^{\prime\prime}}

\def\beq{\begin{equation}}
\def\eeq{\end{equation}}

\newcommand{\beqa}{\begin{eqnarray}}
\newcommand{\eeqa}{\end{eqnarray}}

\let\a=\alpha   \let\b=\beta      
\let\e=\epsilon         
    \let\k=\kappa  \let\l=\lambda  
                 
\let\s=\sigma  \let\t=\tau      

\let\G=\Gamma  \let\D=\Delta

\def\nn{\nonumber}
\let\a=\alpha   \let\b=\beta      
\let\e=\epsilon         
    \let\k=\kappa  \let\l=\lambda  
                 
\let\s=\sigma  \let\t=\tau      

\let\G=\Gamma  \let\D=\Delta

\def\nn{\nonumber}
\newcommand{\be}{\begin{equation}}
\newcommand{\ee}{\end{equation}}
\newcommand{\bea}{\begin{eqnarray}}
\newcommand{\eea}{\end{eqnarray}}

\newcommand{\eq}[1]{Eq.~(\ref{#1})}

\def\A5{(A_5)_{\rm lat}}
\def\thintablerule{\hrule height0.4pt}
\begin{document}
\leftline{\hfill{\small{WITS-CTP-69}}}
\vskip .6cm
\centerline{\LARGE Flavor Corrections in the Static Potential}
\vskip .5cm
\centerline{\LARGE in Holographic QCD}

\vskip 2 cm
\centerline{\large  Dimitrios Giataganas$^1$ and\, Nikos Irges$^2$}
\vskip 1cm
\centerline{\it 1. National Institute for Theoretical Physics}
\centerline{\it School of Physics and Centre for Theoretical Physics}
\centerline{\it University of the Witwatersrand}
\centerline{\it Wits, 2050, South Africa}
\vskip .5cm
\centerline{\it 2. Department of Physics}
\centerline{\it National Technical University of Athens}
\centerline{\it Zografou Campus, GR-15780 Athens, Greece}

\centerline{\it e-mail: dimitrios.giataganas@wits.ac.za, irges@mail.ntua.gr}
\vskip 2.2 true cm
\thintablerule
\vskip 2.0ex
\centerline{\bf Abstract}

We examine the static potential in the presence of flavors in the perturbative backreacted D4/D8 system from localized D8 branes, focusing in particular on the Sakai-Sugimoto model.
For the case of long strings we find the flavor corrections to the static potential which are of exponential form.
We then investigate shorter Wilson loops and express their energy analytically in terms of the lengths of two neighboring Wilson loops.
Moreover, we find that beyond a certain scale the static force in the backreacted background   is reduced compared to one in the probe limit, as expected
due to screening effects. We also compare the string world-sheets in the two backgrounds and find how they get modified by the backreaction.
Our results are supported by numerical computations as well.
Finally we discuss our results in comparison with the lattice data and comment on the issue of physical scales
which seem to lie at the heart of the reason that obstructs our model at this level to fully describe QCD.

\vskip 1.0ex\noindent
\vskip 2.0ex
\thintablerule

\vskip-0.2cm
\newpage

\section{Introduction}

The Wilson Loop (WL) is a an object that contains fundamental information about a
gauge theory at a non-perturbative level.
Its ground state gives us the static potential, i.e. the interaction between two infinitely heavy quarks.
For $SU(N_c)$, the pure gauge theory at zero temperature has a unique, confined phase
with a static potential of the form
\be
E = {\rm const.} + \s L + \frac{c_{1}^{\rm gauge}}{L} + \cdots \label{SP}
\ee
Each of the above terms needs a separate discussion.
The constant is non-physical. Typically one considers the static force $dE/dL$ to get rid of it.
The string tension $\s$ for small $N_c$, can be computed in lattice Monte Carlo simulations as the slope of the
static potential, as a function of the static quark separation $L$. It has dimension of
$length^{-2}$ and for that reason it is usually expressed in terms of the lattice spacing
or better in terms of some other physical dimensionful scale.
Of special importance is the Coulomb like term because its coefficient is dimensionless.
In the ultraviolet (UV), that is at small $L$, it is related to the asymptotically free coupling of the gauge theory and evolves
according to the well known, perturbative renormalization group equations. Towards the infrared, it stops running roughly around
a particular scale, the "Sommer scale" and from there on it assumes a constant value.
This constant value has been predicted to be $c_1^{\rm gauge}=-(d-2) \pi /24$ in $d$-dimensions and for any $N_c$ \cite{LW1}.
The prediction derives from a picture which has a non-critical bosonic string
connecting the two static quarks and the (Gaussian) fluctuations
of the associated world-sheet in a flat transverse space, give rise to this "L\"uscher term".
Even more impressively the value was successfully confirmed in a Monte Carlo simulation \cite{LW2}.
Finally the dots in \eq{SP} stand for unknown corrections, at least from the point of view of the lattice.

Things get more involved when dynamical quarks are added to the gauge theory. To begin,
the string tension can be measured straightforwardly
via Monte Carlo simulations, as in the pure gauge case.
On the other hand, to our knowledge there is no string derived prediction for the flavor correction to the L\"uscher term up to date, at least for QCD.
As far as the lattice is concerned
the only estimate of a correction to the QCD static potential which could perhaps support the effort to extract such a prediction
from string theory is a WL analysis with $N_f=2$ dynamical flavors that appeared recently \cite{Alpha}.
The result of this study is that the value of the Coulomb term coefficient $c_1$ is
consistent with the prediction of the Cornell model \cite{Cornell} $c_1\simeq -0.52$, which means that the $N_f=2$ flavors approximately
double the pure gauge value. What is not clear at all is if this correction,
which we call from now on $c_1^{\rm fl}$, is purely geometrical like it is in the pure gauge theory or dynamical.
From the field theory point of view nothing ensures that a
stringy description even exists in QCD in the presence of dynamical flavors.
We obviously suspect that there may be such a description and we study it in this paper.

Regarding the higher order corrections in \eq{SP}, they are even harder to probe in lattice QCD unless perhaps an
independent input, for what kind of terms one should expect is found.
String theory offers us the possibility to compute such corrections, at least to the pure gauge static potential.
Already in flat space, an infinite number of terms contribute
corrections that come as inverse powers of $L$. The form of this polynomial series depends
on the restrictions that one puts on the derivative expansion of the world-sheet action.
The most general expansion is obtained when only Lorentz invariance constrains the world-sheet action.
If on the top of it one assumes modular invariance, one obtains a restriction of the above series.

The question that naturally arises is which is the string that corresponds to the QCD WL.
The exact answer to this question is not known but we do know a class of especially promising approximations
via special, curved holographic backgrounds using the AdS/CFT correspondence \cite{Maldacena98,Witten98}.
The first such computation was that of the static potential in the ${\cal N}=4$ $SU(N_c)$ SYM, in the limit
where $\l = g_{\rm YM}^2N_c$ is large and fixed and $N_c\to \infty$.
In this setup, the WL is the boundary of a world-sheet that extends in the bulk of a curved
($AdS_5$) space-time. The static potential was derived from the minimization of the world-sheet surface with the appropriate boundary conditions
and yielded a Coulomb-like potential \cite{MalWil,Reywl}, as expected from a conformally invariant gauge theory.

By now there is a good understanding of many extensions of the original Maldacena's proposal. The AdS/CFT correspondence has been
extended to theories with less or zero supersymmetries, as the $\beta$ deformed ones\footnote{The Wilson loop for the $\beta$ deformed theories at least for large $N$, is expected to be similar to one of the $\cN=4$ SYM theory \cite{Giataganasbeta}.} and the quiver theories with the Sasaki-Einstein gravity
duals. On the other hand there have been found also several confining backgrounds at zero and finite temperature, aiming to provide the ground
to understand better QCD. One important development in this direction was the understanding that fields which transform in the fundamental
representation as the QCD quarks do, can be introduced by adding appropriate flavor branes in the gravity dual background \cite{Karchflavor} . Usually
this is done in the probe limit where the backreaction of the flavor branes is neglected, and by studying the worldvolume of these branes one can observe
in the gravity dual background  many physically expected phenomena and to compute for example the meson spectrum\footnote{See \cite{Erdmengerreview} for
a review on the AdS/CFT with flavors.}. Nevertheless, in the quenched approximation the mesons are affected from the glueballs but not the other way around.
So ideally one would like to consider the backreaction of the flavor branes to the original background, in order to obtain more realistic models. The equations
needed to be solved for the derivation of the backreacted backgrounds are in general second order, non-linear partial differential equations, which include
Dirac delta functions due to the localization of the flavor branes. To solve these equations it is a challenging problem. On the other hand one can consider an
approximate approach, and allow a smearing of the flavor branes instead of a localization. This approximation simplifies the differential equations, but
the smearing breaks a $U(N_f)$ flavor group to $U(1)^{N_f}$ and hence theories with $U(N_f)$ flavor groups can not be examined exactly\footnote{See \cite{Nunezreview} for a review on the backreaction of smeared flavored branes.}.

In this paper,  motivated from the previous discussion about the lattice approach and the difficulties encountered in AdS/CFT away from the probe limit,
we study how the static heavy Q\={Q} potential is modified in holographic QCD when the backreaction of the flavor branes is taken into account. For this reason
we use the Sakai-Sugimoto \cite{SS} model which is the closest one so far to the holographic description of large $N_c$ QCD. Taking advantage of the fact
that the backreaction of the localized branes has been found near the tip of the geometry for the D8 and \={D}8 branes placed on the antipodal points of the
compactified circle \cite{Sonnenback}, we argue that the leading corrections to the static quark potential for the particular WL we consider come from this region.
Doing that we achieve to study analytically as well as numerically the flavor corrections on the static potential and the static force. The corrections to the potential in the non-flavored
background for very long strings  are expected to be exponential \cite{Sonnenlec}, and we show that the flavor corrections in this limit are also exponential
although with different factors. As a result, for such strings, we do not find any correction to the string tension or to the L\"uscher term.
In order to compute the static potential of short strings we perform a series expansion which converges to
a string length of order one. We compute the static potential there and we find flavor corrections to the string tension
but no corrections in inverse powers in the string length, beyond those that come from
the $\a'$ expansion. In several cases we compute explicitly all the interesting coefficients appearing in the static potential.
We perform a first comparison of the static force in the holographic approach with lattice QCD data.
A qualitative similarity in their behavior is observed but as we will demonstrate, a quantitative
comparative study is subtle.

The paper is organized as follows. In section 2 we review the Witten-Sakai-Sugimoto background (WSS)
and the localized backreacted solution we are interested in.
In section 3, we first derive the flavor corrections to the energy in the limit of large distance
between the Q\={Q} and find them to be exponential as in the probe approximation but with different factors.
Then we present a method which allows us to compute short an intermediate distance corrections.
The numerical analysis is included in section 4 where we explain how we perform it in order to be valid for our
approximation and we present several results for
the static potential, the color force and the string worldsheet comparison in the two backgrounds.
Finally, we summarize and discuss our results in the last section and include two appendices to support the analysis of the main text.

\section{The general setup}
\subsection{The Witten background for $D=4$}

The unflavored  version of the WSS model, is similar to the background derived by Witten \cite{Wittenss} and consists of $N_c$ D4 color branes,
where one spatial direction is compactified on $S^1$ with radius $\rho$. The submanifold spanned by the radial coordinates $u$ and $x_4$ has a
topology of a cigar which ends at $u=u_k$, where to avoid the singularity at the tip, one requires that
\bea
x^4\sim x^4+ 2\pi \rho ~,\qquad \rho:=\frac{2}{3}
\frac{R^{3/2}}{u_{k}^{1/2}} ~ .
\eea
The background dual to the 5-d pure Yang-Mills theory is the D4 brane metric solution \cite{Wittenss}
\bea
&&ds^2=\left(\frac{u}{R}\right)^{3/2}
\left(\eta_{\mu\nu}dx^\mu dx^\nu+f(u)dx_4^{2}\right)
+\left(\frac{R}{u}\right)^{3/2}
\left(\frac{du^2}{f(u)}+u^2 d\Omega_4^2\right),\label{Wmetric}
\\\nn
&&e^\phi= g_s \left(\frac{u}{R}\right)^{3/4},
~F_4=dC_3=\frac{2\pi N_c}{V_4}~\epsilon_4 \ ,
~f(u)=1-\frac{u_k^3}{u^3} \ ,~R^3=\pi g_s N_c l_s^3 ~,
\label{D4ss}
\eea
where $\mu=0,..,3$ and
$d\Omega_4^2$, $\epsilon_4$ and $V_4=8\pi^2/3$
are the line element, the volume form and the volume of a unit $S^4$,
respectively. Also, the string length is $l_s = \sqrt{\a'}$.
The couplings, satisfy the following relations:
\bea\nn
&& g_5^2=(2\pi)^2 g_s l_s~,\quad g_4^2=
\frac{g_5^2}{2\pi \rho}=3\sqrt{\pi}\left(\frac{g_s u_k}{N_c l_s}\right)^{1/2},\quad \l_5=g_5^2 N_c
\eea
Clearly, when $\rho\to \infty$ we reach the (UV) limit where the dual gauge theory is 4+1 dimensional and
in the opposite (IR) limit we expect to see an effective 3+1 dimensional theory, possibly with some memory of its
five-dimensional origin. It can be shown \cite{Sonnen1,stringtension} that in the IR the static potential,
consistently with dimensional reduction, develops a string tension equal to
\be
\s^{(0)} = \frac{1}{2\pi\a'} \left(\frac{u_k}{R}\right)^{3/2}\, .
\ee
We will repeat and extend the proof of this later. It was found that when a fair number of higher derivative terms are taken into account in the world-sheet action,
the L\"{u}scher term retains its universality but the string tension does get modified \cite{Aha1}. We review the results regarding $\a'$  corrections in Appendix B.

In the IR limit the static potential is found to be
\be
E_{\rm IR}={\rm const.} + \s^{(1)} L -\frac{\pi}{12L} +{\cal O}(L^{-3})\, ,
\ee
where $\s^{(1)}$ is the  renormalized string tension and the Coulomb like correction is the leading correction in an $\a'$ expansion, the above mentioned L\"uscher term.

The computation of the static potential in the UV is left for section 3, along with a possible alternative approach towards a comparison with lattice gauge theory. From \eq{d5} we find
a 5-d Coulomb potential.
Thus we arrive at a picture of the phase diagram of the Witten model where
the 5-dimensional $SU(N_c)$ gauge theory passes, without crossing a sharp phase transition, from
a Coulomb phase to a 4-dimensional confined phase, as the theory gets dimensionally reduced when
$\rho\to 0$. The theory never looses its 5-dimensional memory since all light states are proportional to
$1/\rho$ and $\rho$ can never really be zero.
In addition, as noticed in \cite{susymeson}, it does not loose its supersymmetric origin either, since there are
fermionic mesons originating from the world-sheet fermions that are also set by the same scale.
We will not compute here flavor corrections in the UV limit, since their immediate importance is questionable.

The puzzle that arises then from all this is that the correction to the Coulomb term that the lattice sees \cite{Alpha} and which is
of order $c_1^{\rm fl}\sim{\cal O}(-1/4)$ can not be explained by Witten's background even if we include
$\a'$ corrections.
Moreover, adding probe flavor D8 branes can not improve the situation because in this limit
the background, and therefore the WL, is unaffected. The only left over possibility we see is to consider the backreaction
of the flavor branes and recompute the WL in the corrected background.

\subsection{The WSS background with backreaction}

The Witten-Sakai-Sugimoto model is the extension of the model  of the previous section with a $U(N_f)\times U(N_f)$
chiral flavor symmetry which is spontaneously broken.
In order to realize the flavor symmetry, one introduces $N_f$ D8 branes and $N_f$ \={D}8 branes
transverse to $x_4$ and localized \cite{SS}. 
In the probe approximation,
where $N_c\gg N_f$, the effect of the flavor branes on the geometry is negligible.
The gravity approximation is valid when $\l_5\gg \rho$. The dual theory in this limit is a $4+1$ dimensional maximally supersymmetric theory
with gauge group $SU(N_c)$, compactified on a circle
(with supersymmetry breaking, antiperiodic boundary conditions on the $S^1$ for the fermions)
and coupled to $N_f$ left handed fermions and $N_f$ right handed fermions in the fundamental
representation of $SU(N_c)$. At energies lower than $M_{KK}=1/\rho$ the dual gauge theory is effectively four dimensional confining, with mass gap of order $M_{KK}$.
At the opposite limit $\l_5\ll \rho$, the theory is a $3+1$ dimensional and approaches a pure Yang-Mills theory where the mass gap is exponentially suppressed compared to $1/\rho$.

The D4 branes extend along $x^{\mu}, x_4$ while the D8 branes along the coordinates $x^\mu,~\Omega_4$ and $u$. At the boundary the branes are localized at $x_4=0$ (D8) and at $x_4=L_8$ (\={D}8). As they extend in the bulk they meet, and the chiral symmetry breaks since the group $U(N_f)_L\times U(N_f)_R$ which exists close to the boundary, breaks to the diagonal $U(N_f)$ inside the bulk. By maximizing the flavor branes' distance to  $L_8=\pi \rho$ at the boundary, the corresponding equations force them to meet at the lowest possible radial point $u=u_k$.
Notice that in this setup the constituent mass of the quarks tends to zero.
A gauge theory where the dynamics of glueballs and mesons influence each other
can be obtained holographically by taking into account the backreaction of the flavor branes on the D4 geometry.
As already mentioned, there are several methods to consider this backreaction and take into accounts its effects.
For our purposes we will focus on the localized backreacted flavor branes approach.

To find the backreaction in the case under study one has to consider the D4-D8 system in the Romans massive type IIA supergravity\footnote{Recently it has been argued that in the UV might be impossible to complete the WSS model. In this limit the massive IIA string theory between the D8-branes seems that can not be strongly coupled while remaining weakly curved \cite{JafferisIIA}.}. The corresponding equations for the dilaton and the graviton have been found \cite{Sonnenback}, and as expected they contain delta functions. To solve these equations analytically in the full range of $u$
is a very challenging problem. Instead of that, one can take advantage of the fact that the delta functions are codimension 1 and try to solve the equations perturbatively. The perturbative approach in this case seems to be consistent since the Green's functions are linear with respect to the distance of the brane and hence finite. This perturbative analysis close to $u_k$ has been done in \cite{Sonnenback}.
The small flavor expansion parameter $q_f$ used is defined as
\be
q_f=g_s\k^2_{10} K_8 = \frac{g_sN_f}{4\pi l_s}
\ee
where $K_8$ is proportional to the 8-dim brane tension and the $\k_{10}$ is proportional to the 10-dim Newton gravitational constant.
Notice that $q_f$ has dimension of inverse length.
The solution when $u\rightarrow u_k$ is parametrized as
\bea\label{deff}
ds^2&=&e^{2A_1(u,x_4)}\left(-dt^2+dx_i^2\right)+e^{2A_2(u,x_4)}dx_4^2+e^{2A_3(u,x_4)}du^2+e^{2A_4(u,x_4)}d\Omega_4^2  \\\nn
\phi(u,x_4)&=&\frac{1}{2}\hat{\phi}(u,x_4)+2A_1(u,x_4)+2A_4(u,x_4)~,\quad F_{(4)}=Q_c V_4 \, ,
\eea
where $Q_c=3 R^3/g_s=3 \pi N_c l_s^3$.
The functions in the exponentials are expanded as
\bea
A_i(u,x_4)=A_{u,i}(u)+q_f A_{d,i}(u,x_4)~, \quad \phi(u,x_4)=\phi_u(u)+q_f \phi_d(u,x_4)~,\nn
\eea
where the $u$ subscripted functions are the usual undeformed solutions of the $q_f=0$ equations, i.e. their exponents are equal to the metric elements of \eq{Wmetric} and the index $i$ runs from one to four. The functions determining the backreaction around $u_{k}$ turn out to be
\bea\nn
&&A_{d,1}=\frac{\sqrt{u^3-u_k^3} \sin\left(\frac{x_4}{\rho}\right) \left(u_k^3 \left(7+2 \cos\left(\frac{2 x_4}{\rho}\right) \right)+4 \sin\left(\frac{x_4}{\rho}\right) ^2 u^3\right)}{54 u_k^{7/2}}~,\\
&&A_{d,2}=A_{d,3}=\frac{4}{5} \sqrt{\frac{u^3-u_k^3}{u_k}}  \sin\left(\frac{x_4}{\rho}\right) ~,\quad\phi_d=\frac{1}{3}  \sqrt{\frac{u^3-u_k^3}{u_k}} \sin\left(\frac{x_4}{\rho}\right)
\eea
and the corresponding metric elements
\bea
&&g_{00}=-g_{xx}=-\exp\left(\frac{ q_f \sin\left(\frac{x_4}{\rho}\right) \sqrt{u^3-u_k^3} \left(u_k^3 \left(7+2 \cos\left(\frac{2 x_4}{\rho}\right)\right)+4 \sin\left(\frac{x_4}{\rho}\right)^2 u^3\right)}{27 u_k^{7/2}}\right) \left(\frac{u}{R}\right)^{3/2}~,\nonumber\\
&&g_{44}=\exp\left(q_f \frac{8}{5} \sin\left(\frac{x_4}{\rho}\right) \sqrt{\frac{u^3-u_k^3}{u_k}}\right) \left(1-\frac{u_k^3}{u^3}\right) \left(\frac{u}{R}\right)^{3/2}~,\\\label{metricf}
&&g_{uu}=\exp\left(q_f \frac{8}{5}  \sin\left(\frac{x_4}{\rho}\right) \sqrt{\frac{u^3-u_k^3}{u_k}}\right)\left(1-\frac{u_k^3}{u^3}\right)^{-1} \left(\frac{u}{R}\right)^{-3/2}\, .
\eea
Using this background we examine the flavor effects in the static Q\={Q} potential in the next section.

\section{Static potential in presence of flavors}

As already mentioned, the full backreacted solution is not known and
only the backreaction close to the tip of the cigar has been calculated. However, based on the
fact that the string we are interested in lies almost entirely close to $u_k$ when
$u_0 \simeq u_{K}$, with $u_0$ being the turning point of the string world-sheet, we claim that our results produce in this limit the leading corrections to the static potential.

The open string between the two heavy quarks, is localized in the $x_4$
dimension at an equal distance from the D8 and  \={D8} branes. 
The string is attached to the boundary and hangs inside the bulk down to a minimum value $u_0$.
In our analysis this minimum is chosen to be close to $u_k$ so that most of the string lies
close to the tip of the geometry. It is then natural to think that the leading flavor effects for the
particular WL come from the region of the world-sheet near $u_k$ and we indeed find that this is true.

The flavor backreaction in our approximation should be perturbative
and close to $u_k$. This is the case close to $u_k$, but as we depart from $u_k$ to higher values of $u$ the elements of the backreacted metric \eqref{deff} diverge relatively fast from the undeformed elements.
Although for our solution the important part of the metric is close to the tip of the cigar geometry, we would like to make sure that the string solution we find is valid in the deformed space.
As we will see next, a localized string at $x_4=\pi \rho/2$, solves the undeformed as well as the deformed equations of motion.
Hence to summarize the picture of our string configuration, we  consider a localized string at
$x_4=\pi \rho/2$ in the static gauge where it has a turning point at $u_0\simeq u_k$. It solves the equations of motion in the perturbative background as well as the original background, and it lies almost entirely close to the minimal point $u_0$ where we argue that the flavor effects there will be the dominant ones.

Along these lines we can perform also numerical computations. In numerics we would like to integrate through the whole background. So we need to introduce a 'cut-off' in the deformed metric to eliminate
the flavor effects as the distance from $u_k$ increases and we go outside of the validity of our approximation. One type of such cut-off could be the Heaviside step function, which would eliminate the effects in a non-continuous way. This could create a minor cusp in the
world-sheet solution. Another way for the numerical integration is to use an exponential type 'cut-off'
\be\label{cutoff}
g_{11}=g_{u11} e^{2 q_f A_{d,1} \exp[-\k(u-b)]} ~,
\ee
and so on for the other elements ($g_{u11}$
is the undeformed metric element).
The constant $\k$ defines how fast the flavor effects reduce as the distance from $u_k$ gets bigger.
The parameter $b$ measures from where these flavor effects start to reduce. It is obvious that
$b= u_0\simeq u_k$ in our case. To determine the value of $\k$, we have to set the (small) maximum
distance from $u_k$ within the regime where the perturbative solution is still valid and outside of which the flavor effects should disappear.
It turns out that for the most of the cases $\k=1$ works well. We continue our numerical analysis in section 4.

\subsection{Flavor corrections for long strings}

Here we calculate analytically the leading flavor corrections in the static potential for long strings.
Following the standard method we consider the string ansatz
\be
t=\t,\qquad x_1=\sigma,\qquad x_4=\pi \rho/2,\qquad  u=u(\sigma)~,
\ee
where the equations of motion come from the Nambu-Goto action
\be
S=-\frac{1}{2 \pi \a'}\int d\s d\t \sqrt{-g_{00}\left( g_{11}+g_{44}x_4'{}^2+g_{uu}u'{}^2\right)}~,
\ee
and the metric elements are the ones of \eq{deff}. The equation of motion for $x_4$ is satisfied for
$x_4=\pi \rho/2$, so can proceed with the usual Hamiltonian formalism for the equation of motion of $u$ and using the general relations derived in Appendix A.
From \eq{uprime0} we solve for the constant $c$ which is the Hamiltonian
and a constant of motion, in terms of the turning point $u_0$. The result reads
\bea\label{ccc}
c=\frac{u_0^{3/2}}{R^{3/2}}\exp\left(\frac{ q_f \sqrt{u_0^3-u_k^3} \left(4 u_0^3+5 u_k^3\right)}{27 u_k^{7/2}}\right)~.
\eea
The integral for the string length including the correction due to the backreaction gives
\bea
L=2 \int_{u_0}^\infty du \sqrt{\frac{e^{\frac{ q_f \left(10 \sqrt{u_0^3-u_k^3} \left(4 u_0^3+5 u_k^3\right)+\left(191 u_k^3-20 u^3\right) \sqrt{u^3-u_k^3}\right)}{135 u_k^{7/2}}} R^3 u_0^3}{\left(u_k^3-u^3\right) \left(e^{\frac{2 q_f \sqrt{u_0^3-u_k^3} \left(4 u_0^3+5 u_k^3\right)}{27 u_k^{7/2}}} u_0^3-e^{\frac{2 q_f \sqrt{u^3-u_k^3} \left(5 u_k^3+4 u^3\right)}{27 u_k^{7/2}}} u^3\right)}}~,
\eea
where expanding in $q_f$ and keeping only $\mathcal{O}(q_f)$ terms we obtain
\be\label{ltot1}
L = L_0 + L_f~,
\ee
with
\be
L_0 =  2  \int_{u_0}^\infty du  \sqrt{\frac{R^3 u_0^3}{\left(u^3-u_0^3\right) \left(u^3-u_k^3\right)}}
\ee
being the corresponding integral obtained in the quenched version of the model and
\bea\nn
L_f=&&- q_f (R u_0)^{3/2} \int_{u_0}^\infty du\frac{1}{135 u_k^{7/2} \left(u^3-u_0^3\right)^{3/2}\sqrt{\left(u^3-u_k^3\right)}}\cdot\\\nn
&& \Bigg(\sqrt{u^3-u_k^3}\left(191 u_0^3 u_k^3+60 u^6-20 u^3 u_0^3-141 u^3 u_k^3\right) \\
&&+\sqrt{u_0^3-u_k^3}\left(-50 u^3 u_k^3-40 u^3 u_0^3 \right)\Bigg)
\eea
the correction from the backreaction.

Treating separately $L_0$ and $L_f$, we first perform the expansion $u_0$ around $u_k$
and keep until quadratic terms. Then we integrate the result and by keeping the leading terms at the limit $u_0\rightarrow u_k$ we obtain
\be
L_{0}=\frac{R^{3/2} \left(33-8 \sqrt{3} \pi -48 \log\left(u_0-u_k\right)+24 \log\left(3 u_k^2\right)\right)}{72 \sqrt{u_k}}~,
\ee
where we have also added the constant term
\be
-\frac{R^{3/2}}{\sqrt{3 u_k}}\pi~,
\ee
which comes from the boundary. The leading contribution
from the turning point is a logarithmically divergent term.
Turning to the flavor term $L_f$ we again expand $u_0$ around $u_k$ and obtain the indefinite integral.
By keeping the leading terms in the definite integral we get
\be
L_{f}=-\frac{4 q_f R^{3/2} u_k}{9\sqrt{3} \sqrt{u_0-u_k}}~,\label{flavorL}
\ee
which again gives a divergent leading term when $u_0\to u_k$.

To find the energy in terms of $L$ we need to invert the total length $L$ given from \eq{ltot1} and solve for $u_0$.
It turns out that the result contains the Lambert W function or product logarithm\footnote{The Lambert
W function is a set of functions which defined by solving the equation $z=W(z) e^{W(z)}$, for any function $W$.
This function is also called product logarithm and can not be expressed in terms of elementary functions.}
\be
u_0=u_k+\frac{4 q_f^2 u_k^3}{108 \mbox{LambertW}\left[\pm\frac{q_f u_k}{ 3^{7/4}} e^{-\frac{B}{2}}\right]^2}~,
\ee
where
\be
B=\frac{11}{16}-\frac{\pi }{2 \sqrt{3}}-\frac{3 L \sqrt{u_k}}{2 R^{3/2}}~.
\ee
Although the Lambert W function can not be expressed in terms of elementary functions, it can be Taylor expanded in $q_f$ giving
\be\label{uocor1}
u_0=u_k\left(1+\sqrt{3} e^{B} \right)\pm   2\cdot 3^{-5/4} q_f u_k^2 e^{\frac{B}{2}} ~.
\ee
Using the well known fact that the constant $c$ of \eq{ccc} is the static force, we substitute the above 'plus' sign solution into \eq{ccc} and integrate with respect to $L$ obtaining the energy as
\be\label{efl}
2\pi \a'E= C+\left(\frac{u_k}{R}\right)^{3/2} L -\sqrt{3}u_k e^{B} - \frac{8}{3^{5/4}} q_f u_k^2 e^{\frac{B}{2}}~,
\ee
with $C$ a constant.
From the above relation it is obvious that the correct solution in \eq{uocor1} is the minus sign solution, since this gives the physically expected
predictions and it is in agreement with the numerical results. Moreover the plus sign expression of the \eq{uocor1} gives contradicting  $L(u_0)$ behavior to the \eq{ltot1}.

The 'minus' sign solution does not give any correction in the leading order of $q_f$. Hence the energy reads
\be\label{efl2}
2\pi \a'E= C+\left(\frac{u_k}{R}\right)^{3/2} L -\sqrt{3}u_k e^{B} + \mathcal{O} \left(q_f^2 e^{c B}\right)~,
\ee
where $c$ is a constant and the $q_f^2$ correction term should come with a positive sign in order for the static force to be screened.
We see that the flavor correction to the energy is similar in form to the correction present already in the undeformed background:
they are both $\sim e^{-L}$, although they have different coefficients.
Evidently, at this level of approximations, we do not see the expected from lattice QCD flavor correction $\sim 1/L$.

\subsection{Flavor corrections for shorter strings}

\subsubsection{Probe approximation limit}

We would like to examine also strings of physical length of order one. This could be relevant in the case
where the string breaks due to a vanishing pion mass at a regime where the static energy is around zero.
In this section for convenience we define $y=u/u_0$ and $A=u_k/u_0$.
In the absence of dynamical flavors we have to solve the following two coupled equations:
\be
{\hat L} := \frac{L}{3\rho} = \sqrt{A}\int_1^\infty \frac{dy}{\sqrt{(y^3-A^3)(y^3-1)}}\label{L}
\ee
and
\be
{\hat E} := \frac{2\pi \a'}{u_k} E  =   \frac{2}{A} \left\{\int_1^\infty dy \left[\frac{y^3}{\sqrt{(y^3-A^3)(y^3-1)}}-
\frac{1}{\sqrt{1-\frac{A^3}{y^3}}}\right]-\int_A^1 dy \frac{1}{\sqrt{1-\frac{A^3}{y^3}}}\right\}\; . \label{E}
\ee
In this dimensionless basis, the infinitely long string ($A\rightarrow 1$), has tension ${\hat \s}^{(0)} = 2$.

For $A<1$ we can expand as
\be
\left( 1 - \frac{A^3}{y^3} \right)^{-1/2} = \sum_{k=0}^{\infty} c_k \left(\frac{A^3}{y^3}\right)^k
\ee
with
\be
c_k = \left(^{-1/2}_{\;\;\;\; k}\right)
\ee
the binomial coefficients. Then, using the equation
\bea\nn
&&\int_1^\infty dy \left[\frac{y^{3+\l}}{\sqrt{y^3(y^3-1)} }-m y^\l \right] \sum_{k=0}^{\infty} c_k \left(\frac{A^3}{y^3}\right)^k\\\nn
&&\hspace{6.1cm}=\sum_{k=0}^\infty c_k A^{3k} \; \left[ \frac{m}{\l + 1-3k} + \frac{\sqrt{\pi}}{3} \frac{\G(k-1/3-\l/3)}{\G(k+1/6-\l/3)} \right]
\eea
we can express the two integrals of ${\hat L}$ and ${\hat E}$ by setting $\l=-3, m=0$ and $\l=0, m=1$ respectively, as
\bea
{\hat L}  &=&  \frac{\sqrt{\pi}}{3}\sum_{k=0}^\infty c_k  \frac{\G(k+2/3)}{\G(k+7/6)} A^{3k+1/2}~,\nn\\
{\hat E}  &=&  {\rm const.}+\frac{2\sqrt{\pi}}{3} \sum_{k=0}^\infty c_k  \; \frac{\G(k-1/3)}{\G(k+1/6)} A^{3k-1}\; .
\eea
From now on we drop the $A$-independent constant from ${\hat E}$ and by defining
\be
a_k:= c_k  \frac{\sqrt{\pi}}{3}\ \frac{\G(k+2/3)}{\G(k+7/6)}, \hskip 1cm b_k:=c_k  \frac{2\sqrt{\pi}}{3}\frac{\G(k-1/3)}{\G(k+1/6)}, \hskip 1cm x:=\sqrt{A}\, ,
\ee
we can rewrite the system as
\bea
{\hat L} &=& \sum_{k=0}^\infty a_k x^{6k+1}~,\nn \\
{\hat E} &=&  \sum_{k=0}^\infty \; b_k x^{6k-2}~. \label{LEs}
\eea
Taking derivatives of both \eq{LEs} with respect to $x$ it is easy to show that
\be
2\pi \a' \frac{d E}{dL} = \left(\frac{u_0}{R}\right)^{3/2}\, ,
\ee
a relationship which actually holds more generally (see for example \cite{Ant}).

Next we have to invert the series of ${\hat L}$.
We first compute the derivative
\be
\frac{dx}{d{\hat L}} = \frac{1}{\sum_{k=0}^{\infty}(6k+1) x^{6k} a_k}\, .
\ee
The solution for $x$ can be then written in terms of an expansion around $x=x_c$
\bea
x = d_0 + d_1 {\hat l} + d_2 {\hat l}^2 + d_3 {\hat l}^3 + \cdots~, \qquad
{\hat l} := {\hat L}-{\hat L}_c ~,\label{expansion}
\eea
where ${\hat L}_c$ is the value of ${\hat L}$ at $x=x_c$, that is ${\hat L}_c=\sum_{k=0}^{\infty} x_c^{6k+1} a_k$.
This immediately implies that $d_0 = x_c$.
The $n$'th coefficient $d_n$ is computed by taking $d^n x/dl^n$ and setting $x=x_c$:
\bea
d_1 := \frac{1}{\sum_{k=0}^{\infty} (6k+1) x_c^{6k} a_k}~,\qquad
d_2 := -\frac{1}{2} \frac{\sum_{k=0}^{\infty} 6k (6k+1) x_c^{6k-1} a_k}{\left(\sum_{k=0}^{\infty} (6k+1) x_c^{6k} a_k\right)^3}
\eea
and so forth.
Substituting this solution in the series for ${\hat E}$ one obtains
\be
{\hat E} = \frac{b_0}{(x_c + d_1 {\hat l} + d_2 {\hat l}^2+\cdots)^2} + b_1 (x_c + d_1 {\hat l} + d_2 {\hat l}^2+\cdots)^4+\cdots
\ee
and putting back the units one obtains the expression for $E(L)$.
An observation is that if we expand the above expression in small ${\hat l}$,
the coefficient of the term linear in ${\hat l}$ converges to the infinite string value ${\hat \s}^{(0)}=2$
from below: by summing $10^5$ terms in the series, the coefficient is $1.999772328$.

An interesting first application of the above formulae is when $x_c=0$.
This corresponds to strings of nearly zero length, i.e. to the UV of the gauge theory, that is to its 5-d Coulomb phase.
Then it is easy to see that the leading contribution to the energy is
${\hat E} = \frac{b_0a_0^2}{{\hat L}^2}$ which means that
\be
{\hat E} = -\frac{\pi^{3/2}}{27} \left(\frac{\G(2/3)}{\G(7/6)}\right)^3\frac{1}{\hat L^2}+\cdots\label{UVE}
\ee
or in terms of the dimensionful quantities that
\bea
E &=& -\frac{4}{27}\left(\frac{\sqrt{\pi}\G(2/3)}{\G(7/6)}\right)^3 \frac{R^3}{2\pi\a'}\frac{1}{L^2}\nonumber\\
&=&  -\frac{2}{27}\left(\frac{\sqrt{\pi}\G(2/3)}{\G(7/6)}\right)^3 \frac{g_5^2 N_c}{4\pi^2}\frac{1}{L^2}\, .\label{d5}
\eea
The coefficient $d_5$ of the five-dimensional Coulomb potential
\be
E = - \frac{d_5/a}{L^2}
\ee
was computed in the mean-field expansion of a five-dimensional pure $SU(2)$ lattice gauge theory, near its bulk phase transition \cite{Fc}.
The lattice coupling, which is the only dimensionless parameter on an infinite lattice, is defined as $\beta=\frac{2N_c a}{g_5^2 }$, where $a$ is the lattice spacing.
The difference between this theory and the unflavored Witten model in the UV, is the adjoint scalar field that the latter
has, but this is not expected to affect very strongly the Wilson loop, which is an object constructed out of gauge fields.
The result of the lattice computation was $d_5/a=0.0626$ for $\beta\simeq 2$  \cite{Fc}.
Let us then consider the Witten static potential in the UV \eq{d5} (see also \cite{Sonnen10}) and
identify its Yang-Mills coupling $g_5$ with the corresponding lattice coupling $g_5^2N_c=\frac{2N_c^2}{\beta} a$.
This should be a valid approximation as effects of finite lattice spacing become small
and could allow us to make a first, crude but interesting comparison between two very different computations of analogous quantities.
This leads us to the relation
\be
d_5/a =  \frac{4}{27}\left(\frac{\sqrt{\pi}\G(2/3)}{\G(7/6)}\right)^3 \frac{1}{4\pi^2}\frac{N_c^2}{\beta} = 0.065 \frac{N_c^2}{\beta}\, .
\ee
For $N_c^2\simeq \beta$ there is an impressive agreement with the prediction of the lattice model.
For large $N_c$ however this requires a large $\beta$ that corresponds to the perturbative regime of the five-dimensional gauge theory,
not near the bulk phase transition where $\beta$ is rather small (and where the $SU(2)$ lattice result comes from).
Clearly at the moment no definite quantitative conclusions can be made but the comparison
certainly motivates a generalization of the mean-field lattice computation to $SU(N_c)$
for general $N_c$, where a more direct comparison can be made.

In order to probe strings of finite length however, we need to understand $x_c=1$, its maximum possible value. This corresponds to expanding
around $u_0=u_k$. Notice that in this case
\be
{\hat L}_c = \frac{\sqrt{\pi}}{3}  \sum_{k=0}^{\infty} c_k  \frac{\G(k+2/3)}{\G(k+7/6)} \simeq 0.70125\label{Ec0}
\ee
approximately. This corresponds to a string of physical length $L_c\simeq 1.14$.
This is approximately also the length where the energy is zero, 
so with massless pions in the theory it should be around the string breaking scale. This could be an interesting point to expand around when we add flavors, which we do next.

\subsubsection{Unquenched flavor}

By introducing flavors we write \eq{flavorL} as
\bea
{\hat L}_f = -\frac{2}{9\sqrt{3}}\e_f \sum_{k=0}^{\infty} c_k x^{2k+1}\quad \mbox{with}\quad
\e_f :=  q_f u_k~,
\eea
where $\e_f$ is the dimensionless flavor expansion parameter.
The previous probe relationships generalize to
\bea
{\hat L}_c &=& \sum_{k=0}^{\infty} x_c^{6k+1} a_k - \frac{2\e_f}{9\sqrt{3}} \sum_{k=0}^{\infty} x_c^{2k+1} c_k~, \\\label{ddd1}
d_1 &:=& \frac{1}{\sum_{k=0}^{\infty} (6k+1) x_c^{6k} a_k - \frac{2\e_f}{9\sqrt{3}} \sum_{k=0}^{\infty} (2k+1) x_c^{2k} c_k}~,\\ \nn
d_2 &:=& -\frac{1}{2} \frac{\sum_{k=0}^{\infty} 6k (6k+1) x_c^{6k-1} a_k - \frac{2\e_f}{9\sqrt{3}} \sum_{k=0}^{\infty} 2k (2k+1) x_c^{2k-1} c_k}
{\left(\sum_{k=0}^{\infty} (6k+1) x_c^{6k} a_k- \frac{2\e_f}{9\sqrt{3}} \sum_{k=0}^{\infty} (2k+1) x_c^{2k} c_k\right)^3}~.
\eea
The maximal $L_c$ does not change significantly when $\e_f$ is small.
The energy can be computed from
\be
2\pi \a'\frac{dE}{dL} = \left( \frac{u_0}{R} \right)^{3/2} \left(1+2q_f \left(\frac{1}{6}\sqrt{\frac{u_0^3-u_k^3}{u_k}}
+\frac{2}{27}\frac{(u_0^3-u_k^3)^{3/2}}{u_k^{7/2}}\right)\right).
\ee
We first rewrite it in terms of dimensionless quantities as
\be
\frac{d{\hat E}}{d {\hat L}} =\frac{2}{x^3} \left[ 1 + 2\e_f \left(\frac{1}{6}\left(\frac{1}{x^6}-1\right)^{1/2} + \frac{2}{27} \left(\frac{1}{x^6}-1\right)^{3/2} \right)\right]\label{Eff}
\ee
and then we integrate term by term.
The first term gives
\be
{\hat E}_0 = - \frac{1}{d_1(x_c+d_1{\hat l}+\cdots)^2}.
\ee
We kept $x_c$ general so that we can check against \eq{UVE}. Indeed it is easy to see that for $x_c=0$ the results agree.
We define $d_{10}$ and $d_{1f}$
\be
d_1 =: \frac{1}{d_{10}+\e_f d_{1f}}
\ee
which can be identified in a straightforward way from \eq{ddd1}. Then, performing also the integrals proportional to $\e_f$ in \eq{Eff} and expanding in $\e_f$, to leading order we obtain
\be
{\hat E} =-\frac{d_{10}^3}{(x_c d_{10}+{\hat l})^2}-\e_f\left(\frac{d_{10}^2 d_{11}}{(x_c d_{10}+{\hat l})^2}+\frac{2 d_{10}^2 d_{11} {\hat l}}{(x_c d_{10}+{\hat l})^3} \right)+\cdots \label{Ecf}
\ee
where we can now set $x_c\simeq 1$.
If we expand in small ${\hat l}$ and sum a large number of terms, we
find a string tension whose unflavored value is corrected only by small flavor effects.

Finally we would like to have an improved analytical handle on strings, with ${\cal O}(1) < {\hat L} < \infty$
so we include a higher order correction to the long string expansion of  \eq{ltot1}:
\be
{\hat L}_f^{(2)} = -\frac{\e_f^2}{2} \left(\frac{13 \left(-53+42 \sqrt{3} \pi \right) }{204120}+\frac{z}{108 \sqrt{3}}+\frac{13\text{  }\log(z)}{1620}\right)
\ee
where
\be
 z := \frac{\sqrt{3}A}{1-A} \;\longrightarrow x = \frac{\sqrt{z(z+\sqrt{3})}}{z+\sqrt{3}}
\ee
a parameter that makes sense for $0<z<\infty$.
The length around which we are expanding the string is now
\be
{\hat L}_c = {\hat \a}_2 \log{z_c} -{\hat \a}_3 z_c + {\hat \b}_1 \sqrt{z_c}
\ee
with
\be
{\hat \a}_2 = \frac{1}{3} - \frac{13\e_f^2}{3240}\, , \hskip 1cm {\hat \a}_3 = \frac{\e_f^2}{216\sqrt{3}}\, , \hskip 1cm {\hat \b}_1 = -\frac{2\e_f}{9\cdot 3^{3/4}}
\ee
Clearly, since only in the leading terms we have kept $u_0\ne u_k$,
one should think of $z_c$ being large.
By performing similar steps as above, when ${\hat l}$ is small we obtain
\be
{\hat E} = \left(2+\frac{3\sqrt{3}}{z_c} + \frac{9}{4z_c^2}+\e_f (\frac{2\cdot 3^{3/4}}{3\sqrt{z_c}} + \frac{20\cdot 3^{1/4}}{3z_c^{3/2}}) +\cdots \right) {\hat l} + \cdots \label{Ecf2}
\ee
where ${\hat l} = {\hat L} - {\hat L}_c$.
Using this expansion ${\hat L}_c$ can be lowered significantly but cannot be taken too low because in order to derive
this result, we have set $u_0=u_k$ in subleading terms.
We can see that as $z_c$ increases as we approach the infinite string tension value 2
(with small flavor corrections) from above.

\section{Numerical analysis}

The numerical analysis can be done either by using the method with the cut-off of \eq{cutoff} discussed in the previous section or by introducing a step function which defines the region within the backreacted flavor metric is valid.
However, one should be careful with the values of the step function and especially of the cut-off otherwise the integrals might not converge. Both of the methods give similar results.

By doing the numerical integration for both E and L and using the renormalization scheme of the subtraction of the infinite mass
quark we obtain the results which are presented in Fig. 1. 

The static force between the Q\={Q} is plotted in Fig. 2.\footnote{Alternatively we could have plotted the renormalized charge $\alpha_{Q {\bar Q}}(L)=L^2 F(L)/(4/3)$.} On the left we show the result for the WSS model for $N_f=0$ and $N_f\ne 0$.
On the right we show the result of the lattice QCD analysis for $N_f=0$ and $N_f=2$. The $N_f=0$ lattice data is taken from \cite{NS1} and
the $N_f=2$ data from \cite{Alpha} and \cite{BF1}. In the two sets of data the extrapolated to zero quark mass value $L_0/a=7.05$
has been kept fixed, with $L_0$ the Sommer scale and $a$ the lattice spacing.
In the WSS model we have also kept a scale $L_0$ fixed, however it is not possible to unambiguously
determine its value. In the plot, we have chosen the crossing point of the flavored and non-flavored data
of the WSS model to coincide with the one of the lattice data.
There is an obvious qualitative similarity between the two plots but at this stage it is not clear if
the fixed scales in the two approaches have the same physical meaning.
For large $L/L_0$, where we know the backreaction effects, the static force in the flavored background is less than the one in the non-flavored case. This could be due to the fact that in the presence of flavors, there are  screening effects caused by the virtual quark-pairs produced between the heavy quarks, and reduce the color force. Our interpretation of the plot for smaller values of $L/L_0$ relies on the validity of the backreacted solution in that regime.
\begin{figure}[t]
\centerline{\includegraphics[width=70mm]{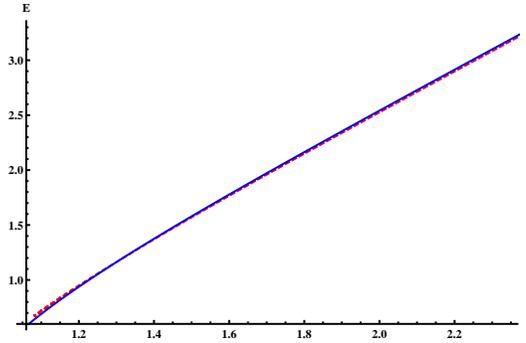}}
\caption{\small{The energy of the Q\={Q} system with numerical integrations in the flavored backreacted background and in the original one. The flavor effects in our analysis come as perturbations so the results are expected to be close at least for large $L$ which correspond to world-sheet turning points close to the $u_k$. We observe that close to $u_k$ the undeformed energy is slightly less than the deformed one.  
The options we follow for all the plots are: The values that are chosen are $q_f u_k = 0.15$, $u_0/u_k\sim 1+5 \cdot10^{-5}$ and increasing slowly close to this region. The flavor backreacted metric is valid only around $u_k$. The energy and the length are divided everywhere by $u_k$. With the red dashed line are the results of the quenched WSS model and with the blue solid line the results of the backreacted WSS model.}}
\end{figure}
\begin{figure}[t]
\centerline{\includegraphics[width=70mm]{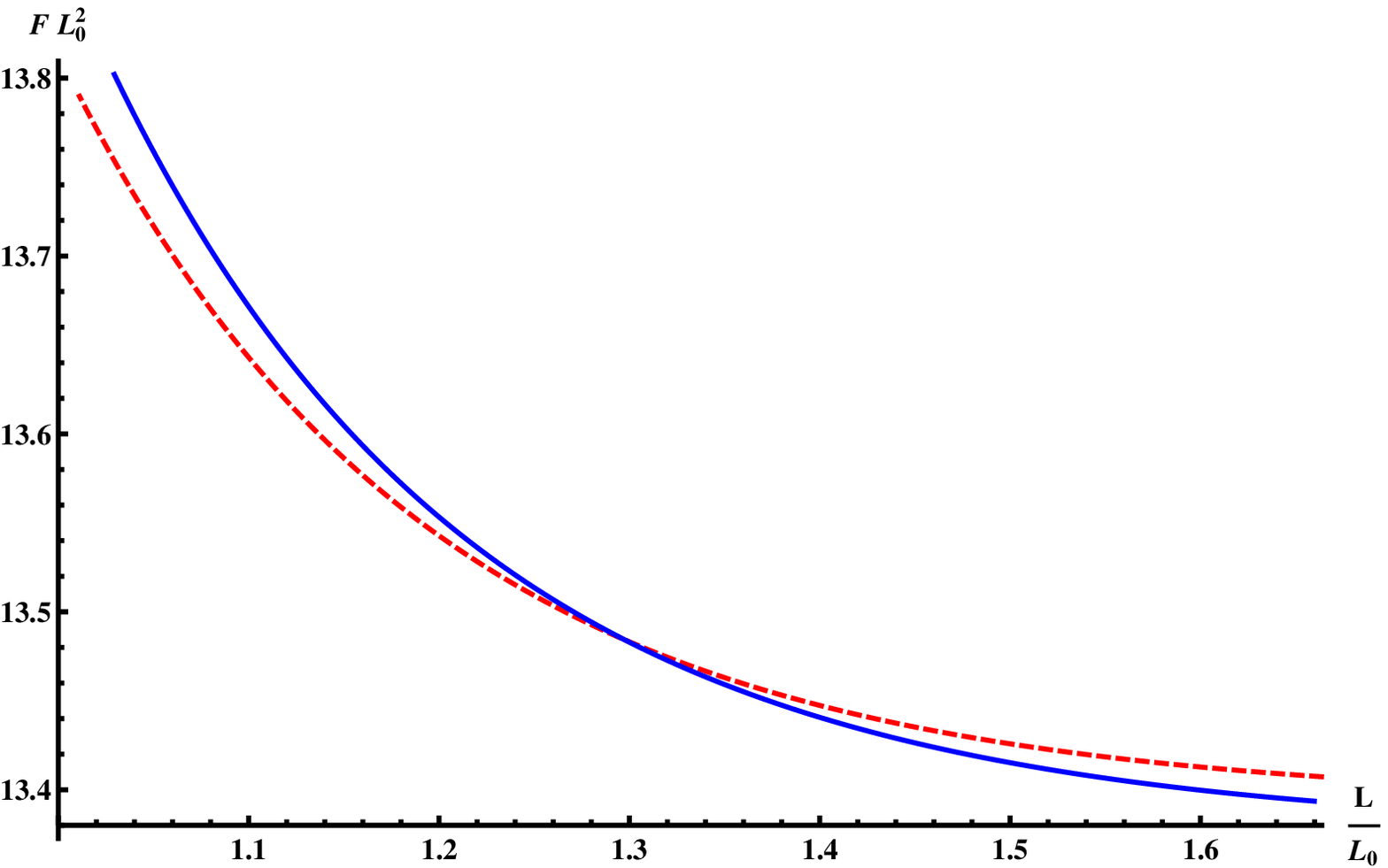}\quad\includegraphics[width=70mm]{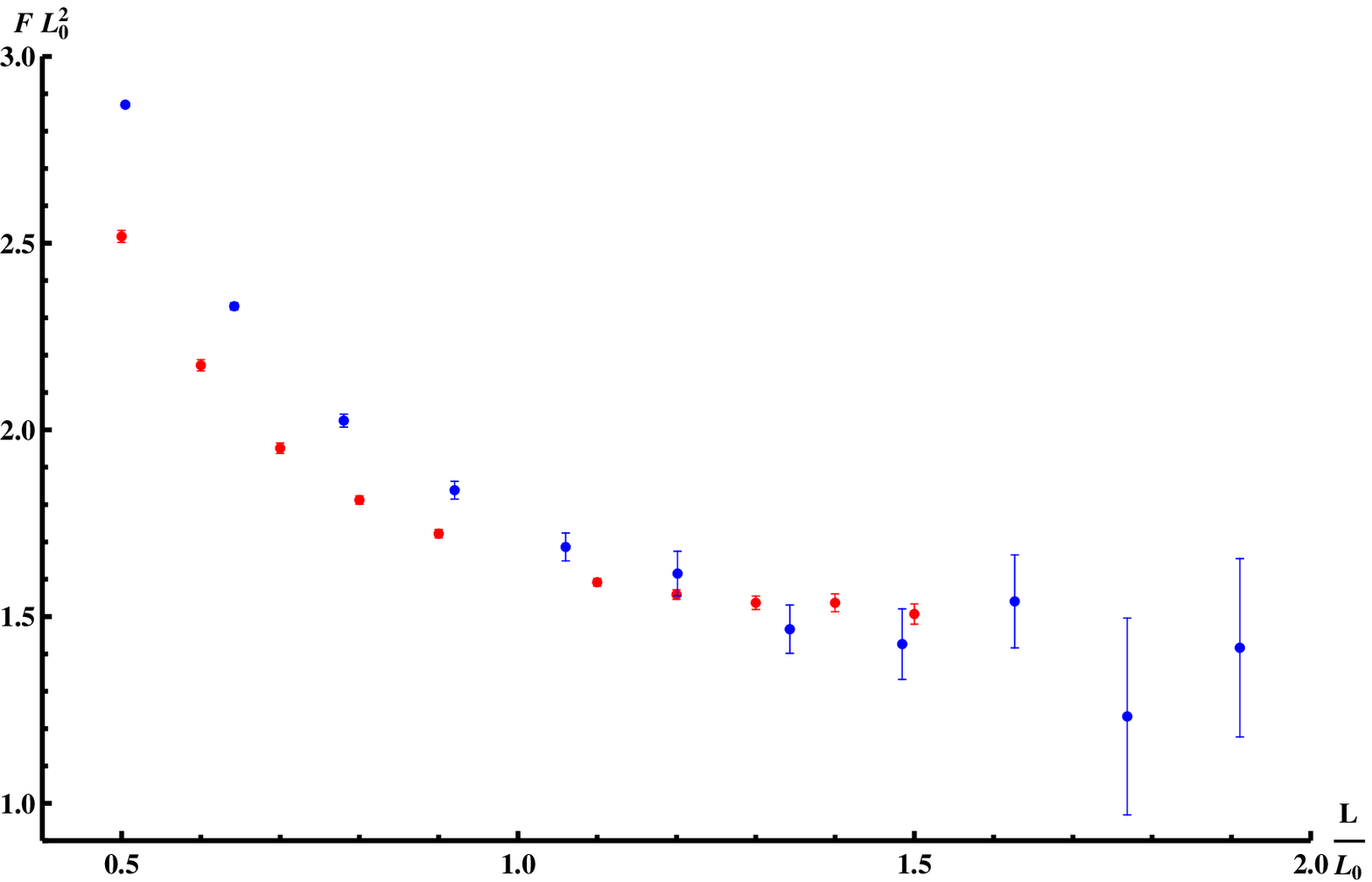}}
\caption{\small{The static Q\={Q} force, as a function of the heavy quark distance. On the left the results obtained from our holographic analysis.
On the right the lattice results for $N_f=0$ and $N_f=2$ taken from \cite{NS1}, \cite{Alpha} and \cite{BF1}. With the red dashed line (red points)
are the results of the non-flavored case and with the blue solid line (blue points) the results of the flavored case of the WSS model (lattice analysis).
In the lattice data $L_0$ is the Sommer scale. In the WSS data it is some fixed scale.}}
\end{figure}

It is interesting to observe that the strings with the same length L, have lower energy in the undeformed background than in the deformed one at least
in the region where $u_0\sim u_k$. Moreover, string world-sheets that are equally radially extended from the boundary have lower energy in the flavored
background, and smaller lengths $L$ are required to reach the minimum distance $u_0$. Hence, to obtain the same Q\={Q} energy in the flavored
and non-flavored background, the string in the flavored case need to be extended less and have smaller length than the one in the original background (Fig. 3).
We expect that these observations carry on in the case of the full backreacted solution of the background.
\begin{figure}[t]
\centerline{\includegraphics[width=70mm]{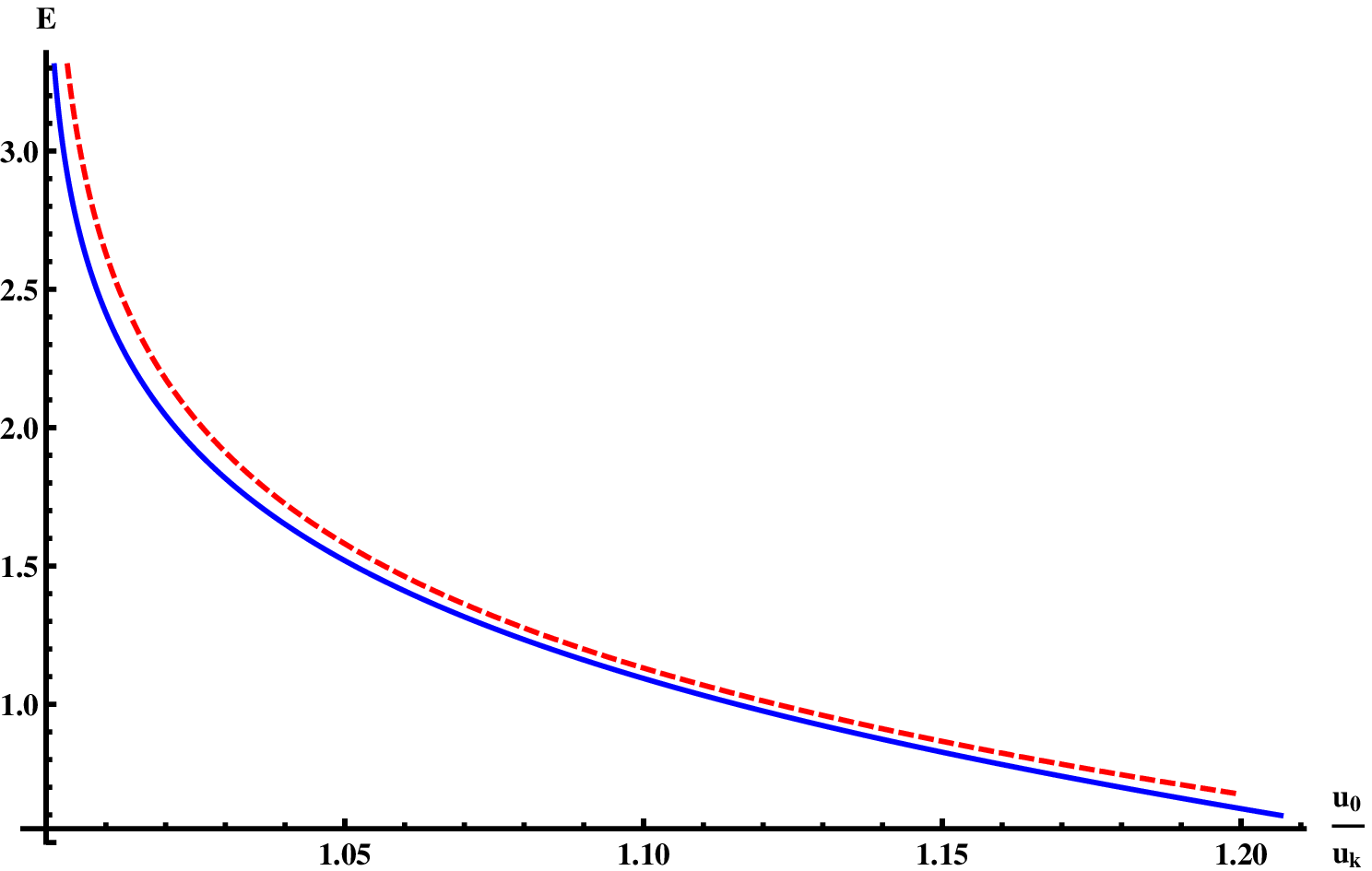}\quad\includegraphics[width=70mm]{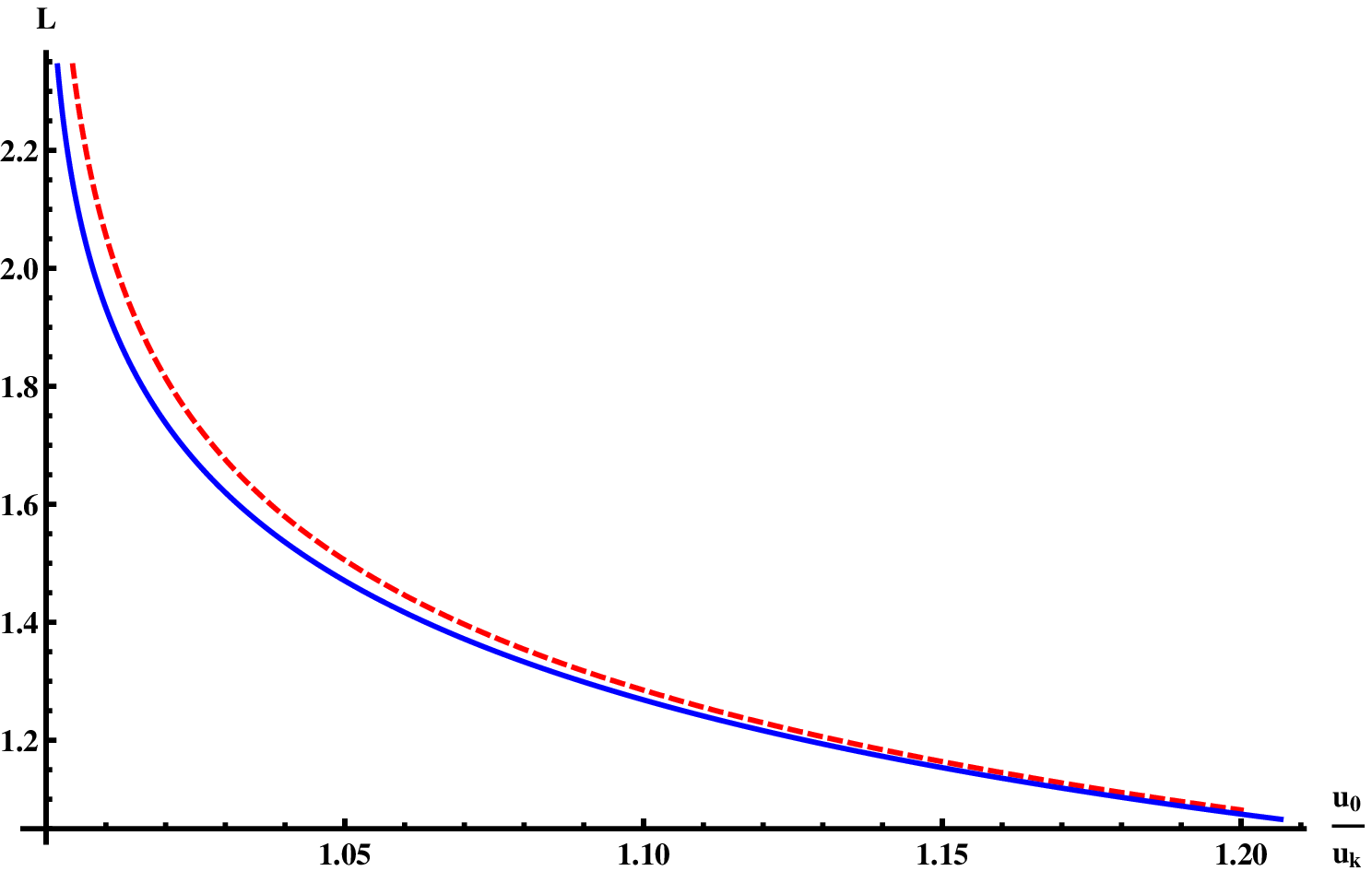}}
\caption{\small{On the left the energy of the string with respect to minimum point $u_0$. On the right the length dependence of the string with respect to $u_0$. The string worldsheet in the flavored background reaches the same energy with the string in the undeformed one for lower values of $u_0$. Furthermore, the string in the flavored background that extended to a minimum point $u_0$ has length $L$ smaller than the string with the same minimum point in the undeformed background.
With the red dashed line are the results of the WSS model in the probe approximation and with the blue solid line the results of the backreacted WSS model.}}
\end{figure}

We observe that since our flavor effects are perturbative the modifications to the backreacted Wilson loop are small.  However there are important differences as the turning point goes far away from $u_k$ and this happens because there our approximation stops to be valid and because of the mass subtraction term which still gets contributions from the region close to $u_k$. When $u_0$ gets outside the region where the flavor effects are taken into account, the first terms of the equation \eq{energt} do not receive any flavor contribution, but the last one coming from the mass of the infinite quark does get important contributions no matter how far from $u_k$ the  $u_0$ is. This situation would be different, if for the regularization scheme the Legendre transform was used. However to use this renormalization to subtract the divergence, the conditions for the background derived in \cite{Giataganasuv} must be satisfied. In our approximation it seems that these are satisfied at least approximately but would be interesting if one could check it for the whole backreacted background.


\section{Discussions and Conclusion}

We have touched upon several problems that obstruct the WSS model from describing
QCD. We have focused on the fundamental properties encrypted in the static potential and static force rather on
shortcomings at the phenomenological level.
We have seen that the regime where the flavor correction to the L\"uscher coefficient
develops that the lattice seems to observe between the Sommer scale and the string breaking scale,
cannot be located. We consider this particular flavor effect in lattice QCD special because
if it admits a string description, it is an order 2 effect in the L\"uscher coefficient.
For that reason, we have scanned analytically via two different expansions
and also numerically the WSS Wilson Loop at essentially all scales. We have seen no sign of
a regime where flavor corrections to the string tension in inverse powers of $L$ develop.
We have argued that $\a'$ corrections are unlikely to change this picture.
According to this discussion the perturbative effects on the
backreacted background may not be enough, and possibly a full backreacted solution is needed
in order to understand via holography the phenomena observed in lattice QCD.
In case these problems persist even if a fully backreacted background is found,
it would mean that a more drastic modification of the WSS model is necessary.

The color force beyond a certain scale reduces in the presence of flavors due to screening effects caused
by the virtual quark-pairs produced between the heavy quarks. We have found this behavior for large distances between the quarks, where the
validity of our approximation is best. This has been also noticed in the backreacted D3/D7
model where the screening effects result to a change of slope in the Regge trajectories \cite{Vamanregge}. A naive comparison between lattice data and our results shows qualitative agreement as depicted in Fig.2. Nevertheless, at this point one should be aware of two sources
of uncertainty. One is the validity of the range of the backreacted solution which could result in the
low $L/L_0$ part of the static potential not to be trusted. The other is related to the identification of
$L_0$ as the Sommer scale in the WSS model.

A full understanding of screening in the WSS model and more generally of its physical scales is subtle.
The origin of the subtlety is an obstruction in the precise determination of a physical length scale:
in the unflavored model the absence of a Sommer scale and in the flavored case in addition a difficulty in determining
the string breaking scale.
The former obstruction is due to the fact that as the theory is followed from the IR towards the UV
one does not seem to pass through a regime where it is 4-d and asymptotically free, nor crosses a
first order phase transition. Instead it passes smoothly into a 5-d Coulomb phase, as \eq{d5} indicates.
There is not much to say about this problem as at present seems to be an intrinsic property of the WSS model.
We could only suggest a possible way out which involves the construction of a background dual
to an anisotropic 5-d gauge theory as opposed to an isotropic one, as it is the case here.
The discussion of the latter obstruction is also subtle.
To begin, for the same reason as in the unflavored case there is no Sommer scale.
One would hope that the backreacted background contains information about the breaking of the string,
which would give us an alternative way to extract a physical scale.
In the Veneziano limit of the MN background\footnote{The analysis in the D3/D7 system can be found at \cite{d3d7}.} system where backreaction of the smeared flavor branes
have been considered \cite{ParedesWL}, it has been
noticed that the backreacted background contains the information of the string breaking and
as the number of flavor branes is reduced compared to color branes,
the length where the string breaking happens increases for massive dynamical quarks. Hence, one could think that in our setup,
because the backreaction comes only as perturbation to the original
background, string breaking happens at very large $L$ close to infinity, where due to the constraints in numerical integrations we could not see it in the plots.
 In our setup however the D8 and \={D}8 branes are placed antidiametrically on the circle. Therefore the constituent quark mass is zero since it is
given by the energy of a string extended from the minimal position of the background to the minimal radial position of the corresponding
flavor branes.
Moreover,  the pions in our model are massless and one would rather physically expect that string breaking happens at short lengths, since a very small amount
of energy is needed to create pions. According to this argument and our results, the string would break in this case around $L\sim 1$.
Hence there seem to be two opposite ranges on the scale of string breaking and in our approximation it seems not to be possible to get a conclusive answer just from our model.

As a result, the length scale around which one should expand could possibly be far from the domain of validity of
the backreacted background. This is not necessarily the case of course, but it is not possible to
perform a reliable quantitative treatment until a better understood version of the model with massive pions is constructed.
In principle it is possible to incorporate some mechanism that gives vev for the chiral condensate and the quark mass \cite{tachyon}.
There are two main difficulties that we see at present with this.
One is that we do not know of a formalism fully under control for $N_f>1$ and the second is that
one should take into account the resulting modifications in the backreaction.

Concerning the behavior of the string world-sheets we have found that the strings with the same length $L$, have lower energy in the undeformed background than in the deformed one at least in the region where $u_0\sim u_k$. Moreover, string world-sheets that are equally radially extended from the boundary have lower energy in the flavored background, and need for smaller lengths $L$ to reach the minimum distance $u_0$. As a result to have same Q\={Q} energy in the flavored and non flavored background, the flavored string is extended less and have smaller length than the one in the original background. These observations should carry on in the full backreacted  solution too.

A possible interesting analysis of the static potential could be done also away of $u_k$, for short string which stay close to the boundary. There the perturbative backreaction is known \cite{Sonnenback}, but the analysis seems to be doable only at numerical level. The backreaction at this range should be also similar to the NJL dual gravity theory \cite{KutHarv}, and that is an additional interesting point. However, the most challenging problem would be to obtain the full backreacted solution of the WSS model and even better by including the bifundamental tachyon field in order to have massive pions. This could be done either by considering the localized branes or by the smearing techniques. If the full solutions were obtained, for small backreaction (small number of flavored branes compared to the color ones) our results should be reproduced. Additionally, in this background one should be able to study explicitly the string breaking and make a comparison with the lattice data. The extension to the finite temperature backreacted background is also of big importance especially for quark-gluon plasma physics since the effect of the dynamical quarks to the plasma physics should be taken into account.


\textbf{Acknowledgements:}

We would like to thank E. Kiritsis, F. Knechtli and R. Sommer for illuminating discussions
and A. Cotrone for useful comments.
N.I. is partially supported by the NTUA support program for basic research ${\rm \Pi EBE}$ 2009 and ${\rm \Pi EBE}$ 2010. The
research of D.G. is supported by a Claude Leon postdoctoral fellowship.

\startappendix

\Appendix{Strings in general weakly coupled backgrounds}

In this section we derive the turning point of world-sheet of a string in static gauge in weakly coupled backgrounds and their energy.
The confining string in the QCD backgrounds, is a special case of the strings we examine here. The following equations have been derived in various forms in several in other papers too \cite{Sonnenlec,stringtension}.
By writing the metric of the space as
\be
ds^2=g_{00}d\t^2+g_{ii}dx_i^2+g_{uu}du^2
\ee
we choose  the static gauge for the string
\be
x_0=\t\qquad\mbox{and}\qquad x_1=\sigma,
\ee
which is extended in the radial direction too, so $u=u(\s)$.
Then the induced metric $G_{\alpha \beta}=g_{MN}\partial_\alpha X^M \partial_\beta X^N$ for our world-sheet read:
\bea
G_{00}=g_{00},\qquad G_{11}=g_{11}+g_{uu}u'^2.
\eea
Supposing that we are working in Lorentzian signature the Nambu-Goto action is\footnote{In the case of Euclidean signature, the formulas change with a minus sign wherever the $g_{00}$ element is.}
\bea
S=\frac{1}{2\pi\a'}\int d\s d\t \sqrt{- g_{00}(g_{11}+g_{uu}u'^2)}=:\frac{1}{2\pi\a'}\int d\s d\t \sqrt{D}~.
\eea
The Hamiltonian then is equal to
\be
H=\frac{ g_{00} g_{11}}{\sqrt{D}}
\ee
and is a constant of motion. Setting it equal to $c$ we can solve for $u'$ and get the turning point equation
\be\label{uprime0}
u'=\pm\sqrt{-\frac{(g_{00} g_{11}+ c^2)g_{11}}{ c^2 g_{uu}}}~,
\ee
which is solved for
\be\label{tpsol}
g_{uu}^{-1}=0~,\qquad\mbox{or}\qquad g_{11}=0~,\qquad\mbox{or}\qquad g_{00}g_{11}=- c^2~.
\ee
The above equations, normally the last one, specify how deep the world-sheet goes into the bulk, and we call the value of the turning $u_{0}$.

The length\footnote{The limits of the length integral depend on where we choose as the starting point measuring $L$. When the string in the boundary extends from $-L/2$ to $L/2$, then the corresponding solution of the $u'$ is positive for $(0,L/2)$ since the turning point corresponds to $L=0$. When the string extends from $(0,L)$ the $u'$ in $(0,L/2)$ is negative. In any case the final result in the definite integral is \eq{LL}.} of the two endpoints of the string on the brane is given by
\be\label{LL}
L=2\int_{\infty}^{u_{0}}\frac{du}{u'}=2\int_{u_{0}}^{\infty}  du \sqrt{\frac{- g_{uu} c^2}{(g_{00}g_{11}+ c^2)g_{11}}}~.
\ee
Moreover, the energy of the string using as renormalization method the mass subtraction of the two free quarks is
\bea\label{energt}
2\pi\a' E&=&2\left(\int_{u_{0}}^{\infty} d\s \cL -\int_{u_{0}}^{\infty} du \sqrt{ g_{00}g_{uu}}\right)\nonumber\\
&=& c L +2\left[  \int_{u_{0}}^{\infty} du \sqrt{- g_{uu}g_{00}}\left(\sqrt{1+\frac{c^2}{g_{11}g_{00}}}-1\right)- \int_{u_{k}}^{u_0} du \sqrt{-g_{00}g_{uu}}\right]~.
\eea
Notice that we already used that fact that the world-sheet is symmetric with respect to turning point $U_0$ and hence the RHS of the above equation are already multiplied by two.

\Appendix{The D4 Wilson loop in the IR}

In the static gauge $X_0=\s$, $X_1=\tau$ the metric \eq{Wmetric} can be brought in the form
\be
ds^2 = \left(1+\frac{m_b^2}{2T}(Y\cdot Y)\right) dX\cdot dX + dY\cdot dY +R^2 d\Omega_4^2
\ee
with $dX\cdot dX=dX_2^2+dX_3^2$, $Y\cdot Y = Y_4^2+Y_5^2$, $T=2\pi\a' \s^{(0)}$ and where
\be
m_b=\frac{1}{\rho}\,
\ee
is essentially the "Kaluza-Klein" scale governing the lightest glueball.
This immediately suggests us that only the transverse $X_{2,3}$ fluctuations are massless and thus that the
induced L\"uscher term in the static potential is likely to be the same as that of a
four-dimensional non-critical bosonic string in a flat background,
$c_1^{\rm gauge}=-\pi/12$.
Next, we would like to actually perform the integral
\be
e^{-S_{\rm eff}[X]} = \int dY e^{-S[X,Y]}\, .
\ee
The result is \cite{Aha1}
\bea
S_{\rm eff}[X] &=& \s^{(1)} \int_{\cal M} d^2\s \left[ 1+\frac{1}{2}(\partial_\a X)\cdot (\partial_\a X)+\cdots\right]\nonumber\\
&+& \sqrt{\s^{(1)}}\int_{\partial {\cal M}} d\s^0 \left[\mu_B + b_1 (\partial_1 X)\cdot (\partial_1 X)+\cdots \right]
\eea
where the dots stand for the world-sheet fermions and higher derivative terms and
\be
\s^{(1)} = \s^{(0)} (1 + \D T)
\ee
is the corrected string tension renormalized by
\be
\D T = \frac{1}{8\pi}  \left[\sum_f\frac{m_f^2}{T}\log \frac{m_f^2}{T} - \sum_b\frac{m_b^2}{T}\log \frac{m_b^2}{T} \right] \, .
\ee
The $m_f$ are world-sheet fermion masses.
There is a wave function renormalization involved to bring the result in the above form, which is finite
provided that the condition $\sum_b m_b^2 = \sum_f m_f^2$ holds.
This condition holds for most known holographic backgrounds, including
the one under study.
The constant can also be computed
\be
\mu_B = -\frac{1}{8} \left[ \sum_a m_a - \sum _{a'} m_{a'} \right]\, ,
\ee
where the index $a$ counts Dirichlet and $a'$ Neumann boundary conditions.
Finally, the coefficient in the boundary terms vanishes: $b_1=0$.
All the above implies that after integrating out also the $X$-fluctuations,
the static potential of the Witten (and in fact of a fairly large number of holographic) model in the IR can be summarized as
\be
E_{\rm IR}={\rm const.} + \s^{(1)} L -\frac{\pi}{12L} +{\cal O}(L^{-3})\, .
\ee
Evidently while the string tension is renormalized, the L\"uscher coefficient retains its universality.
The coefficient $-\pi/12$ assumes that the
perturbatively zero modes on the $S^4$ \cite{co11} pick up a mass at the non-perturbative level \cite{Aha1}.

The perturbative backreaction used in this paper, if it has any effect on the L\"scher term it will
probably give a small mass to the massless transverse bosonic fluctuations and perhaps correct by a small
amount the already massive fields. Thus, the corresponding term could change into
a term of the Yukawa form
\be
\frac{(-\pi /12) e^{-mL}}{L}
\ee
with $m$ the small mass due to the backreaction. Clearly, this will be a small effect.\footnote{The sigma model one loop corrections in $AdS_5\times S^5$ have been studied in \cite{for1}.}


\end{document}